\documentclass[12pt]{article}
\usepackage{latexsym,epsfig,amssymb, amsmath, cite, mcite}

\usepackage[vcentermath]{youngtab}          

\newcommand{\cN}{{\cal N}}

\def\bfone{\relax{\rm 1\kern-.35em 1}}

\newcommand{\be}{\begin{equation}}
\newcommand{\ee}{\end{equation}}
\newcommand{\ben}{\begin{displaymath}}
\newcommand{\een}{\end{displaymath}}
\newcommand{\bea}{\begin{eqnarray}}
\newcommand{\eea}{\end{eqnarray}}

\newcommand{\eins}{\mbox{$1 \hspace{-1.0mm} \text{l}$}}
\newcommand{\bean}{\begin{eqnarray*}}
\newcommand{\eean}{\end{eqnarray*}}
\newcommand{\bbZ}{\mathbb{Z}}
\newcommand{\bbQ}{\mathbb{Q}}
\newcommand{\bbP}{\mathbb{P}}
\newcommand{\Tr}{{\rm Tr}}
\newcommand{\cR}{\mathcal{R}}
\newcommand{\cI}{\mathcal{I}}
\newcommand{\cM}{\mathcal{M}}

\newcommand{\diag}{\rm diag}

\newenvironment{matr}[1]
{\left[ \begin{array}{{#1}}}{\end{array} \right]}

\makeatletter \@addtoreset{equation}{section} \makeatother

\begin{document}

\begin{titlepage}

\begin{flushright}
\small ~ \\
\end{flushright}

\bigskip

\begin{center}

\vskip 2cm

{\LARGE \bf Minimal Stability in \\[.2cm]   Maximal Supergravity} \\

\vskip 1.0cm

{\bf Andrea Borghese$^1$, Rom\'{a}n Linares$^2$, Diederik Roest$^1$}\\

\vskip 0.5cm

{\em $^1$ Centre for Theoretical Physics,\\
University of Groningen, \\
Nijenborgh 4, 9747 AG Groningen, The Netherlands\\
{\small {\tt \{ a.borghese, d.roest \} @rug.nl}}} \\

\vskip 0.3cm

{\em $^2$ Departamento de F\'{\i}sica, \\
Universidad Aut\'onoma Metropolitana Iztapalapa,\\
San Rafael Atlixco 186, C.P. 09340, M\'exico D.F., M\'exico\\
{\small {\tt lirr@xanum.uam.mx}}}

\end{center}

\vskip 2cm

\begin{center} {\bf ABSTRACT}\\[3ex]

\begin{minipage}{13cm}
\small

Recently, it has been shown that maximal supergravity allows for non-supersymmetric AdS critical points that are perturbatively stable. We investigate this phenomenon of stability without supersymmetry from the sGoldstino point of view. In particular, we calculate the projection of the mass matrix onto the sGoldstino directions, and derive the necessary conditions for stability. Indeed we find a narrow window allowing for stable SUSY breaking points. As a by-product of our analysis, we find that it seems impossible to perturb supersymmetric critical points into non-supersymmetric ones: there is a minimal amount of SUSY breaking in maximal supergravity.

\end{minipage}

\end{center}

\vspace{2cm}

\vfill

\end{titlepage}


\tableofcontents

\section{Introduction}

More than thirty years after its inception, maximal supergravity continues to be a fascinating theoretical edifice. Issues of current interest include its possible finiteness as a theory of quantum gravity, its connection to the worldvolume theory of M2-branes and its possible holographic applications in condensed matter systems. In the latter two of these developments, a subgroup of the $E_{7(7)}$ theory has been elevated to a local symmetry, leading to maximal gauged supergravity. The prime example is the $SO(8)$ gauging \cite{N=8a, *N=8b}, which arises from an $S^7$-reduction of $D=11$ supergravity.

Perturbative physics arises as an expansion around a chosen vacuum. A proper understanding of the vacuum structure of a given theory is therefore of the utmost importance. For the $SO(8)$ theory, the vacuum structure has been investigated in ever increasing detail over the last decades. The critical points that preserve an $SU(3)$ gauge group have been exhaustively classified using analytic properties \cite{Warnerold1}. Recently this method has been complemented with numerical techniques, with which it is possible to look for critical points preserving a smaller fraction of the gauge group. Indeed, in addition to the seven classic examples preserving at least $SU(3)$ or $SO(4)$, a large number with gauge groups consisting of zero, one or two factors of $U(1)$ have now been found  (in addition to one exceptional $SO(4)'$ case) \cite{Fischbacher, *Fischbacher2}. All critical points are Anti-de Sitter, and lie strictly below the maximally supersymmetric one preserving $SO(8)$ at the origin. Furthermore, a number of new properties concerning the spectra of two of these critical points have come to light. 

First of all, the mass spectrum was calculated for the critical point that preserves an $SU(4)^-$ subgroup  \cite{Warnernew1}. This serves, in a truncated setting, as the endpoints of the flow that emerges in the zero temperature limit of the holographic superconductor solution of \cite{Gauntlett}. However, it turns out that in the full $\cN=8$ theory, the spectrum has an average mass of\footnote{In all of this paper but figure 1, we will express all masses in units of the scalar potential $V$. This is related to the (A)dS length $L$ via $|V| L^2 = 3$.} $m^2 / |V| = - 3/ 10$ but includes twenty tachyonic scalars with a mass $m^2 / |V| = -1$. As this is below the Breitenlohner-Freedman bound  \cite{BF} of $m^2 /|V| \geq -3/4$, this signals an instability. This is perhaps not very surprising, as this critical point breaks supersymmetry completely. Indeed, the expectation has been for a long time that within maximal supergravity, there are no non-supersymmetric and nevertheless stable critical points.

The second new result involves the mass spectrum of the $SO(4)$ invariant critical point. Again, this critical point breaks all SUSY and hence was expected to be perturtbatively unstable. However, an explicit calculation showed that this is not the case \cite{Warnernew2}. While the average mass in this case amounts to $m^2 / |V| = 6 /35$, the lowest scalar masses are again twentyfold and come in at $m^2 / |V| = - 4/7$. Contrary to expectation, and thanks to the BF bound, this critical point is therefore perturbatively stable. Therefore it turns out to be possible to attain stability without supersymmetry, also in maximal supergravity.

A similar phenomenon has subsequently been found in a different setting. Most theories other than $SO(8)$ have only been investigated in the origin. Examples include the unstable De Sitter critical points of $SO(4,4)$ and $SO(5,3)$ \cite{HW84}. However, an exhaustive classification of all $SO(3)$ critical points has also been performed in the half-maximal theories that arise in IIA geometric compactifications with O6/D6 sources \cite{DGR}. Surprisingly, out of a myriad of possibilities,  it turns out that only a single theory has such critical points, of which there are four distinct ones. Moreover, this compactification requires a vanishing net O6/D6 charge, and hence can be embedded in the maximal theory. The full mass spectrum of one of the non-supersymmetric critical points turns out to have an average value of $m^2 / |V| = 16/5$ and the lowest masses are given at $m^2 / |V| = 0$. Therefore this critical point is also perturbatively stable, in this case even without the BF bound, again while breaking all supersymmetry. 

In light of the renewed interest in the vacua structure of these different maximal supergravity theories, it seems to be of interest to analyse in full generality to what extent one can make statements concerning stability. For instance, given these newfound non-supersymmetric vacua with perturbative stability, a natural question concerns their multitude in the landscape of critical points. Are they rare occurances, as perhaps suggested by history, or are they in fact very commonplace but have we been looking in the wrong corner of the landscape so far? In other words, how easy is it to preserve stability while breaking supersymmetry?

In the best possible scenario, it would be feasible to analyse the stability of all supersymmetry breaking critical points for all maximal supergravities in one fell swoop, including all possible gaugings and hence scalar potentials. However, as can be expected, the possibilities to analyse vacuum stability are rather limited in such a general setting. The full mass matrix is 70-dimensional, and hence cannot be diagonalised directly. One will have to consider projections onto lower-dimensional subspaces. In this way one can derive necessary (but not necessarily sufficient) conditions for stability. The simplest possibility that comes to mind is the trace of the mass matrix. As we will show, this average mass does not teach one much. Beyond this, one needs specific directions onto which to project the mass matrix. 

For configurations that break supersymmetry (even if only partly), the sGoldstini furnish specific directions in scalar space. sGoldstini are the scalar partners of the dilatini that are absorbed by the eight gravitini in the process of supersymmetry breaking. Their relevance in distinguishing unstable directions in $\cN = 1$ supergravity was pointed out in \cite{GR1, *GR2, *GR3}, while this has been generalised to theories with extended supersymmetry in \cite{Louis, *Jacot, BR}. In the present paper we will follow the analysis of \cite{BR} and apply this to the most general theory of maximal supergravity. 

A sneak preview of the main results of this paper includes the following highlights. Indeed the sGoldstino directions allow for a measure of the ratio between the number of stable and unstable critical points. We find that a very small fraction of parameter space allows for stable sGoldstino directions. As this is only a necessary condition, the actual set of stable, non-supersymmetric critical points is expected to be even smaller. This holds true for AdS, Minkowski and dS critical points, and explains the notion of minimal stability of the title. Furthermore, we find that there is a finite gap between the maximally supersymmetric critical point and all non-supersymmetric critical points. This stems from the impossibility to introduce arbitrary supersymmetry-breaking effects in maximal supergravity, unlike e.g.~F- and D-terms in $\cN = 1$. So-called quadratic constraints prevent one from doing this in $\cN = 8$. Thus it seems that there is a minimal amount of supersymmetry breaking that is needed to deform a supersymmetric critical point to a non-supersymmetric one. The notion of approximate supersymmetry ceases to exist for critical points of maximal supergravity.

Upon completion of this paper, a number of preprints \cite{McAllister, Chen, DI} were submitted that address related issues. In the context of random $\cN = 1$ supergravity, it was found \cite{McAllister} that the likelihood of stable De Sitter vacua is exponentially supressed as a function of the number of moduli fields. Furthermore, the most promising places to look at are approximately-supersymmetric critical points. Our results can be seen as an analytic\footnote{Note that a random sample of $\cN = 8$ theories is non-trivial to define due to the quadratic constraints.} $\cN = 8$ counterpart of this. As there are 70 scalar fields in maximal supergravity, and moreover the notion of approximate supersymmetry might not exist, it seems doubtful that there could be a stable dS vacua at all. This is corroborated by our results. Furthermore, new critical points of $\cN = 8$ gauged supergravity and their mass spectra were presented in \cite{DI}, employing a search method along the lines of \cite{DGR}. It would be very interested to relate these results to ours.

This paper is organised as follows. General background on maximal supergravities is introduced in section 2, while section 3 discussed the sGoldstino projection and resulting mass values. In section 4 we go through a number of explicit examples of critical points, while section 5 contains a general analysis of the constraints implied by the sGoldstino mass. Finally, we conclude in section 6.

\section{Maximal supergravity}

Maximal supergravity is the unique theory in four dimensions with eight real supersymmetries. Its field content consists of a graviton, eight gravitini, 28 vectors, 56 dilatini and 70 scalar fields. Together these form the supergravity multiplet, while there are no matter multiplets for this theory. The symmetries of the theory include a global $E_{7(7)}$, which is not a symmetry of the Lagrangian and can only be realised on the equations of motion, however. In addition, there is a local $SU(8)$ R-symmetry. 

In addition to the field content, also the interactions of maximal supergravity are constrained to a high degree by supersymmetry. In particular, in the ungauged theory, all interactions are determined: the theory is unique. The only freedom to introduce interactions in the theory is to consider gauged supergravity instead. The vectors allow one to gauge a 28-dimensional subgroup of $E_{7(7)}$. This gauging completely determines the additional interactions, and furthermore introduces non-zero masses in the theory. 

A convenient framework to describe the most general maximal gauged supergravity is the so-called embedding tensor \cite{Samtleben, *Samtleben2} . It determines which $E_{7(7)}$ generators are gauged by which vector. As a consequence, the embedding tensor takes values in the
tensor product of the fundamental $\bf 56$, in which the electric and magnetic components of the tensor transform, with the adjoint $\bf 133$, in which the $E_{7(7)}$ generators transform. This product is given by
 \begin{align} \label{Embedding tensor irreps}
   {\bf 56} \times {\bf 133} = {\bf 56} + {\bf 912} + {\bf 6480} \,.
 \end{align}
Supersymmetry implies that only gaugings corresponding to the $\bf 912$ are consistent. It can be represented by a constant tensor $X_{MNP}$, where $M$ corresponds to the fundamental $\bf 56$, subject to the linear conditions
 \begin{align} \label{Linear constraints in E7}
 X_{M[NP]} = 0 \, , \qquad X_{(MNP)} = 0 \, , \qquad X_{MN}{}^{N} = X_{MN}{}^{M} = 0 \, ,
 \end{align}
that restrict $X_{MNP}$ to the $\bf 912$. In addition consistency of the gauging requires the embedding tensor to satisfy the following quadratic constraints, living in the $\bf 133$ and $\bf 8645$, respectively:
 \begin{align} \label{Quadratic constraints in E7}
 [X_{M},\, X_{N}] = - \, X_{MN}{}^{P} \, X_{P} \, , \qquad X_{MN}{}^{P} \equiv [X_{M}]_{N}{}^{P} 
 \end{align}
The $\bf 912$ irrep of the embedding tensor completely determines the form of the theory, and in particular of its mass spectrum. We will in particular be interested in the scalar mass spectrum in critical points of the scalar potential.

The scalar fields span the coset $E_{7(7)} / SU(8)$. By virtue of the scalar manifold being a homogeneous space, one can always employ the non-compact generators of the isometry group to transform any critical point to the origin. The remaining symmetry is then given by the isotropy group. In other words, we sacrifice the $E_{7(7)}$ symmetry and remain with its compact subgroup $SU(8)$ as a symmetry. The advantage of this procedure is that one is expanding all scalar quantities around the origin, where this expansion takes a particularly nice form. We would like to stress that this does not consistute a loss of generality: given any critical point of any maximal gauged supergravity, this can always be brought to the origin with an $E_{7(7)}$ transformation, where our analysis applies.

After going to the origin of moduli space, the decomposion of a number of relevant $E_{7(7)}$ irreps into $SU(8)$ reads
 \begin{align} \label{SU8 Branching embedding tensor}
  {\bf 56} & \rightarrow {\bf 28} + {\bf c.c.}~\,, \quad
  {\bf 133}  \rightarrow {\bf 63} + {\bf 70} \,,\quad
  {\bf 912}  \rightarrow {\bf 36} + {\bf 420} + {\bf c.c.}~\,, 
 \end{align}

The decomposition of the $\bf 133$ corresponds to the isometries of the scalar manifold: the $\bf 63$ are its compact isometries while the $\bf 70$ are its non-compact isometries. Only the latter correspond to physical scalars, which will be parametrised by
 \begin{align} \label{Self duality on scalars}
   \phi_{ijkl} = \tfrac{1}{4!} \, \epsilon_{ijklmnpq} \phi^{mnpq} \,.
 \end{align}
The scalar kinetic terms are canonically normalised and in the origin read
 \begin{align} \label{Scalar kinetic in the origin}
  {\mathcal L} = -\tfrac{1}{96} \, (\partial_\mu \phi_{ijkl})^2 \,.
 \end{align}
The index $i$ denotes the fundamental $\bf 8$ of $SU(8)$. Indices are raised and lowered by complex conjugation. In addition, for four antisymmetrised indices, one can impose the self-duality relation (\ref{Self duality on scalars}) involving the Levi-Civita tensor. This corresponds to a reality condition on the $\bf 70$ irrep, which therefore splits up in real (anti-)self-dual irreps ${\bf 70}_\pm$. In this notation, the scalars take values in the ${\bf 70}_+$.

The decomposition of the $\bf 912$ corresponds to the embedding tensor, which parametrises all the gaugings and hence all the masses of this theory. We will denote the two resulting $SU(8)$ tensors as
 \begin{align} \label{SU8 embedding tensor}
    {\bf 36}: ~~ A_1 \equiv A_{ij} \, , \qquad 
    {\bf 420}: ~~ A_2 \equiv A^{i}{}_{jkl} \,,
 \end{align}
where the former is symmetric and the latter is anti-symmetric and traceless. Their role is as follows:
 \begin{itemize}
 \item
  $A_1$ is the scale of supersymmetric AdS,
 \item
  $A_2$ is the order parameter of SUSY breaking.
 \end{itemize}
However, in contrast to $\cN = 1$ theories, these different tensors are not independent. Rather they have to satisfy the following quadratic constraints:
  \begin{align} \label{Quadratic constraints I}
& 0 = 9 \, A_{r}{}^{stm} A^{r}{}_{sti} - A_{i}{}^{rst} A^{m}{}_{rst} - \delta_{i}^{m} \, |A_2|^{2}  \, , \\
& 0 =  3 \, A_{r}{}^{stm} A^{r}{}_{sti} - A_{i}{}^{rst} A^{m}{}_{rst} + 12 \, A_{ir} A^{mr} - \tfrac{1}{4} \, \delta_{i}^{m} \, |A_2|^{2} - \tfrac{3}{2} \, \delta_{i}^{m} \, |A_1|^{2} \,, \notag \\
& 0 = A^{i}{}_{jv[m} \, A^{v}{}_{npq]} + A_{jv} \, \delta^{i}_{[m} \, A^{v}{}_{npq]} - A_{j[m} \, A^{i}{}_{npq]} + \notag \\
& \qquad + \tfrac{1}{4!} \, \epsilon_{mnpqrstu} \, \left( A_{j}{}^{ivr} \, A_{v}{}^{stu} + A^{iv} \, \delta_{j}^{r} \, A_{v}{}^{stu} - A^{ir} \, A_{j}{}^{stu} \right) \, , \notag \\ 
& 0 = A_{i}{}^{rsm} A^{n}{}_{jrs} - A_{j}{}^{rsn} A^{m}{}_{irs} + 4 \, A^{(m}{}_{ijr} A^{n)r} - 4 \, A_{(i}{}^{mnr} A_{j)r} + \nonumber \\
 & \qquad - \tfrac{1}{8} \, \delta_{i}^{n} \left( A_{r}{}^{stm} A^{r}{}_{stj} - A_{j}{}^{rst} A^{m}{}_{rst} \right) + \tfrac{1}{8} \, \delta_{j}^{m} \left( A_{r}{}^{stn} A^{r}{}_{sti} - A_{i}{}^{rst} A^{n}{}_{rst} \right)  \,, \notag \\
 & 0 = A_{r}{}^{mnp} A^{r}{}_{ijk} - 9 \, A_{[i}{}^{r[mn} A^{p]}{}_{jk]r} - 9 \, \delta_{[i}^{[m} \, A_{j}{}^{rs|n} A^{p]}{}_{k]rs}  - 9 \, \delta_{[ij}^{[mn} \, A_{r}{}^{p]st} A^{r}{}_{k]st} + \delta_{ijk}^{mnp} \, |A_{2}|^{2}  \, , \notag
\end{align}
corresponding to the ${\bf 63}, {\bf 63}, {\bf 70}_- + {\bf 378} + {\bf 3584}, {\bf 945} + {\bf \overline{945}}$ and $ {\bf 2352}$ irreps, respectively. From the third equation we can extract the pure ${\bf 70_{-}}$ and ${\bf 378}$ irreps, taking the trace and antisymmetrising. We get\footnote{Note that there is a little typo in the expression for the $\bf 378$ in the third line of (D.4) in \cite{LeDiffon}. This can be corrected by requiring this equation to live in the right irrep.}
\begin{align} \label{Quadratic constraints II}
& 0 = A^{r}{}_{[ijk} A_{l]r} - \tfrac{3}{4} \, A^{r}{}_{s[ij} A^{s}{}_{r|kl]} - \tfrac{1}{4!} \, \epsilon_{ijklmnpq} \left( A_{r}{}^{mnp} A^{qr} - \tfrac{3}{4} \, A_{r}{}^{smn} A_{s}{}^{rpq} \right)  \,, \\
& 0 = \tfrac{3}{4} \, A^{r}{}_{ijk} A_{lr} + \tfrac{3}{4} \, A^{r}{}_{l[ij} A_{k]r} - \tfrac{1}{4!} \, \epsilon_{mnpqrijk} \, A_{l}{}^{qrs} A_{s}{}^{mnp} - \tfrac{3}{4} \, \tfrac{1}{4!} \, \epsilon_{ijklmnpq} \, A_{r}{}^{spq} A_{s}{}^{rmn} \, . \notag 
\end{align}
All these constraints are required for consistency of the gauging and follow from the the decomposition of (\ref{Quadratic constraints in E7}) with respect to $SU(8)$.

From the general theory of maximal gauged supergravities it follows that the scalar potential in the origin is given by
 \begin{align} \label{Scalar potential}
   V = - \tfrac34 \, |A_1|^{2} + \tfrac{1}{24} \, |A_{2}|^{2} \, .
 \end{align} 
Furthermore, the mass matrix for the 70 real scalar fields reads \cite{LeDiffon}
 \begin{align}
  m^2_{ijkl}{}^{mnpq} = & \; + \delta_{ijkl}^{mnpq} \, \left( \tfrac{5}{24} \, A^{r}{}_{stu} \, A_{r}{}^{stu} - \tfrac{1}{2} \, A_{rs} \, A^{rs} \right) + 6 \, \delta_{[ij}^{[mn} \, \left( A_{k}{}^{rs |p} \, A^{q]}{}_{l]rs} - \tfrac{1}{4} \, A_{r}{}^{s |pq]} \, A^{r}{}_{s|kl]} \right) \nonumber \\
& \; - \tfrac{2}{3} \, A_{[i}{}^{[mnp} \, A^{q]}{}_{jkl]} \, .
 \label{Mass-matrix}
 \end{align}
In general this will be a scalar dependent quantity, but being in the origin it is completely determined by the two embedding tensor components. The mass averaged over all 70 scalars corresponds to the properly normalised trace of this mass matrix and is given by
 \begin{align} \label{Average mass}
  \text{Tr}\{m^{2} \} = - \tfrac{1}{2} \, |A_1|^{2} + \tfrac{1}{20} \, |A_{2}|^{2} \, .
 \end{align}
Finally, the requirement of the origin to be a critical point translates into the additional quadratic constraint
 \begin{align} \label{Field equations}
 A^{r}{}_{[ijk} A_{l]r} + \tfrac{3}{4} \, A^{r}{}_{s[ij} A^{s}{}_{r|kl]} = - \tfrac{1}{4!} \, \epsilon_{ijklmnpq} \left( A_{r}{}^{mnp} A^{qr} + \tfrac{3}{4} \, A_{r}{}^{smn} A_{s}{}^{rpq} \right) \,,
 \end{align}
which correspond to the field equations of the scalars and therefore lives in the ${\bf 70}_+$.

Note that for supersymmetric critical points, for which $A_2$ vanishes, all scalar masses coincide and are given by $m^2 / |V| = -2/3$, corresponding to the discrete unitary irreducible irrep of the AdS$_4$ isometry group \cite{Nicolai}. Indeed, this is the mass spectrum of the $SO(8)$ invariant critical point, for which $A_1$ can be chosen to be proportional to the identity. This is in fact the only $\cN = 8$ critical point, as we will now prove.

Specialising to $A_2 = 0$, (\ref{Quadratic constraints I}) and (\ref{Quadratic constraints II}) reduce to a single constraint 
\begin{align}
A_{ik} \, A^{kj} = \tfrac{1}{8} \, \delta_{i}^{j} \, |A_{1}|^{2}  \quad \Longrightarrow \qquad \cM_{ik} \cM^{kj} = \tfrac{1}{8} \, \delta_{i}^{j} \, ,
\end{align}
with $\cM_{ij} = A_{ij} / \sqrt{|A_1|^{2}}$. $\cM$ is a complex symmetric matrix which can thus be written as $\cM = \cR + i \, \cI$ with $\cR$ and $\cI$ symmetric and real. Taking the product with the complex conjugate we get
\begin{align}
\cM \cM^{*} = \cR \cR + i \, [\cI,\, \cR] + \cI \, \cI = \tfrac{1}{8} \, \eins_{8} \, .
\end{align}
As the right hand side is a real quantity, we must necessarily have $[\cI,\, \cR] = 0$. Thus $\cR$ and $\cI$ can be diagonalised simultaneously. Hence $\cM$ can be brought in the form 
\begin{align}
  \cM = \diag\{ e^{i \, \theta_1},\, e^{i \, \theta_2},\, e^{i \, \theta_3},\, e^{i \, \theta_4},\, e^{i \, \theta_5},\, e^{i \, \theta_6},\, e^{i \, \theta_7},\, e^{i \, \theta_8} \} \, .
\end{align}
Using an $SU(8)$ transformation we can eliminate 7 phases, making $\cM$ equal to the identity modulo an overall phase. This proves that the only possible gauging with $A_2=0$ and therefore $\cN = 8$ is the $SO(8)$ gauging.

 The goal of this paper is to investigate the mass spectrum for non-supersymmetric critical points, which are more difficult to analyse due to the complicated $A_2$ contributions to (\ref{Mass-matrix}).

\section{sGoldstini directions}

As there are eight gravitini becoming massive in the process of supersymmetry breaking, there are also eight Goldstini. Their supersymmetric partners, the sGoldstini, are therefore 64-fold and will be denoted by $V^{rs}_{ijkl}$. Their explicit form is completely determined by the order parameter of supersymmetry breaking, and in components is given by
 \begin{align}
   V^{rs}_{ijkl} = \delta^{r}_{[i} \, A^{s}_{\; \, jkl]} \,.
 \end{align}
These split up in the symmetric $\bf 36$ and the antisymmetric $\bf 28$ irreps. At any stationary point, in line with the less supersymmetric cases \cite{Louis, *Jacot, BR}, the set of 36 symmetric scalar directions $V^{(ij)}$ correspond to physical sGoldstini scalars, while the set of 28  antisymmetric scalar directions $V^{[ij]}$ are pure gauge. The latter are directions that have been gauged away by the gauging induced by $A_1$ and $A_2$. 

From the above set of scalar directions, one can define the following Hermitian projectors:
 \begin{align}
   \bbP_{ijkl}{}^{mnpq} = V^{(rs)}_{ijkl} \, V_{rs}^{mnpq} \,, \qquad
   \bbQ_{ijkl}{}^{mnpq} = V^{[rs]}_{ijkl} \, V_{rs}^{mnpq} \,.
 \end{align}
Due to the interpretation of the antisymmetric sGoldstini as gauge directions, the mass matrix vanishes after projecting with $\bbQ$, as we will see below. Instead we will be interested in the projection of the mass matrix with $\bbP$, yielding information on the physical scalars.

In order to describe the various quartic contractions of the $A_1$ and $A_2$ tensors that appear when calculating the sGoldstini mass, it will be convenient to introduce the following notation. Different contractions are denoted by 15 real coordinates $\vec x$ and 7 complex coordinates $\vec z$, with explicit components
 \begin{alignat}{2} \label{xz-space}
  x_1 & = |A_{2}|^{2} \, |A_2|^{2} \, , &
  x_2 & = |A_{2}|^{2} \, |A_1|^{2} \, , \nonumber \\
  x_3 & = |A_{1}|^{2} \, |A_1|^{2} \, , &
  x_4 & = A_{ir} A^{mr} \, A_{ms} A^{is} - \tfrac{1}{8} |A_{1}|^{2} \, |A_{1}|^{2} \, ,\nonumber \\
  x_5 & = A_{r}{}^{stm} A^{r}{}_{sti} \, A_{u}{}^{vzi} A^{u}{}_{vzm} \, , &
  x_6 & = A_{r}{}^{stm} A^{r}{}_{sti} \, A_{m}{}^{uvz} A^{i}{}_{uvz} \, , \nonumber \\
  x_7 & = A_{i}{}^{rst} A^{m}{}_{rst} \, A_{m}{}^{uvz} A^{i}{}_{uvz} \, , &
  x_8 & = A_{r}{}^{stm} A^{r}{}_{sti} \, A_{mu} A^{iu} \, , \nonumber \\
  x_9 & = A_{i}{}^{rst} A^{m}{}_{rst} \, A_{mu} A^{iu} \, , &
  x_{10} & = A_{r}{}^{smn} A^{r}{}_{sij} \, A_{t}{}^{uij} A^{t}{}_{umn} \, , \nonumber \\
  x_{11} & = A_{r}{}^{smn} A^{r}{}_{sij} \, A_{m}{}^{tui} A^{j}{}_{ntu} \, , &
  x_{12} & = A_{i}{}^{rsm} A^{n}{}_{jrs} \, A_{m}{}^{tui} A^{j}{}_{ntu} \, , \nonumber \\
  x_{13} & = A_{i}{}^{rsm} A^{n}{}_{jtu} \, A_{m}{}^{tuj} A^{i}{}_{ntu} \, , &
  x_{14} & = A_{i}{}^{rsm} A^{n}{}_{jrs} \, A^{ij} A_{mn} \, , \nonumber \\
  x_{15} & = A_{r}{}^{s[ij} A_{s}{}^{r|kl]} \, A^{t}{}_{uij} A^{u}{}_{tkl} \, , \nonumber & \\
  z_1 & = \tfrac{1}{4!} \, \epsilon_{ijklmnpq} \, A_{r}{}^{ijk} A^{lr} \, A_{s}{}^{mnp} A^{qs} \, , & 
  z_2 & = \tfrac{1}{4!} \, \epsilon_{ijklmnpq} \, A_{r}{}^{sij} A_{s}{}^{rkl} \, A_{t}{}^{umn} A_{u}{}^{tpq} \, , \nonumber \\
  z_3 & = A^{r}{}_{s[ij} A^{s}{}_{r|kl]} \, A_{t}{}^{ijk} A^{lt} \, , &
  z_4 & = \tfrac{1}{4!} \, \epsilon_{ijklmnpq} \, A_{r}{}^{sij} A_{s}{}^{rkl} \, A_{t}{}^{mnp} A^{qt} \, , \nonumber \\
  z_5 & = \tfrac{1}{4!} \, \epsilon_{ijkmnprs} \, A_{t}{}^{ijk} \, A^{vz} A_{v}{}^{mnp} A_{z}{}^{rst} \, , \quad & 
  z_6 & = A_{r}{}^{smn} A^{r}{}_{sij} \, A^{i}{}_{mnt} A^{jt} \, , \nonumber \\
  z_7 & = \tfrac{1}{4!} \, \epsilon_{ijkmnprs} \, A_{v}{}^{zrt} A_{z}{}^{vsu} \, A_{t}{}^{ijk} A_{u}{}^{mnp} \, . & & 
 \end{alignat}
The $\vec x$ and $\vec z$ can be seen as coordinates of a vector space. We will later find that not all domains of this space are admissible for viable gaugings. 

The metastability of any non-supersymmetric critical point of maximal supergravity  can be investigated by considering projections of the mass matrix. In particular, we project the mass matrix using the symmetric sGoldstini scalars:
 \begin{align} \label{sGoldstino mass definition}
  M^2_{\rm sG} = \bbP_{ijkl}{}^{mnpq} \, m^2_{mnpq}{}^{ijkl} \,.
 \end{align}
We will refer to this quantity as the sGoldstino mass, even though it does not necessarily correspond to any of the eigenvalues of the mass matrix (\ref{Mass-matrix}). Instead, it consists of the sum of the diagonal elements of the mass matrix corresponding to the 36 sGoldstini. Expression (\ref{sGoldstino mass definition}) is quartic in the embedding tensor components. In order to obtain a mass we need to normalise it dividing by the trace of the projector $\bbP$. It is given by
\begin{align} \label{sGoldstino length}
\delta_{mnpq}^{ijkl} \, \bbP_{ijkl}{}^{mnpq} = \tfrac{3}{4} \, |A_2|^{2} \, .
\end{align}
Hence we define the properly normalised sGoldstino mass as
\begin{align} \label{Normalised sGoldstino mass}
m_{\rm sG}^{2} =\frac{M_{\rm sG}^{2}}{\frac{3}{4} \, |A_2|^{2}} \, .
\end{align}
Clearly, in order for a critical point to be stable, a necessary condition is that the normalised sGoldstino mass is positive or, in the case of AdS, is above the Breitenlohner-Freedman bound $m^2 = \tfrac34 V$. 

The sGoldstino mass turns out to be given by
 \begin{align} \label{sGoldstino mass}
  M^2_{\rm sG} = \tfrac{17}{96} \, x_1 - \tfrac{3}{8} \, x_2 - \tfrac{3}{8} \, x_5 - \tfrac{5}{16} \, x_6 + \tfrac{1}{48} \, x_7 - \tfrac{9}{16} \, x_{10} + \tfrac{27}{16} \, x_{11} + \tfrac{9}{16} \, x_{13} \, .
 \end{align}
From the above form it is not obvious how one can make definite statements about its value. However, there is a number of constraints on the embedding tensor components that we can use in order to simplify the sGoldstino mass. 

Recall that the $A_1$ and $A_2$ tensors that define the scalar potential and hence the sGoldstino mass are subject to a number of quadratic constraints (\ref{Quadratic constraints I}). In the following we will show a simple example of how these constraints can be used to reduce the number of independent $\vec{x},\, \vec{z}$. Consider the first two expressions of (\ref{Quadratic constraints I}) both living in the $\bf 63$ . Subtracting the second from the first and multiplying the result with its complex conjugate, we get the following relation quartic in $A_1$ and $A_2$
 \begin{align} \label{Example quartic relation I}
0 & = A_{r}{}^{stm} A^{r}{}_{sti} \, A_{u}{}^{vzi} A^{u}{}_{vzm} - \tfrac{1}{8} \, \left( |A_2|^{2} \right)^{2} - 4 \, \left( A_{ir} A^{mr} \, A_{ms} A^{is} - \tfrac{1}{8} |A_{1}|^{2} \, |A_{1}|^{2} \right) \, .
 \end{align}
It is clear that, using the dictionary \eqref{xz-space}, this can be interpreted as a hyperplane in $\vec x$-space. 
 \begin{align} \label{Example quartic relation I}
0 & = x_5 -  \tfrac{1}{8} \, x_1 - 4 \, x_4 \, .
 \end{align}
The full set of such restrictions reads
\begin{align} \label{Quartic relations complete}
\text{from the {\bf 63}} \qquad & \begin{cases} 0 \, = \, x_5 - \tfrac{1}{8} \, x_1 - 4 \, x_4 \, , \\
 0 \, = \, x_6 - \tfrac{1}{8} \, x_1 - 36 \, x_4 \, , \\
 0 \, = \, x_7 - \tfrac{1}{8} \, x_1 - 324 \, x_4 \, , \\
 0 \, = \, x_8 - \tfrac{1}{8} \, x_2 - 2 \, x_4 \, ,  \\
 0 \, = \, x_9 - \tfrac{1}{8} \, x_2 - 18 \, x_4 \, , \end{cases} \nonumber \\
\text{from the {\bf 70}}_{-} \qquad & \begin{cases}  0 \, = \, \tfrac{1}{2} \, x_9 - \tfrac{3}{2} \, x_{14} - (z_1 + \bar{z}_{1}) - \tfrac{3}{4} \, (z_3 + \bar{z}_3 ) + \tfrac{3}{4} \, ( z_4 + \bar{z}_4 ) \, , \\
 0 \, = \, \tfrac{3}{4} \, (z_3 + \bar{z}_3 ) - \tfrac{3}{4} \, ( z_4 + \bar{z}_4 ) + \tfrac{9}{16} \, (z_2 + \bar{z}_2 ) - \tfrac{9}{8} \, x_{15} \, , \end{cases} \nonumber \\
\text{from the {\bf 378}} \qquad & \begin{cases} 0 \, = \, x_9 + x_{14} - \tfrac{4}{3} \, z_5 - z_4 \, , \\
0 \, = \, \tfrac{4}{3} \, \bar{z}_5 + \bar{z}_4 - \tfrac{1}{6} \, x_6 - x_{11} - \tfrac{1}{2} \, x_{13} + \tfrac{3}{4} \, x_{15} \, , \end{cases}  \\
\text{from the {\bf 945}} \qquad & \begin{cases} 0 \, = \, x_{10} - x_{12} - 4 \, z_6 + 4 \,  z_3  - \tfrac{1}{8} \, x_5 + \tfrac{1}{4} \, x_6 - \tfrac{1}{8} \, x_7 \, , \\
0 \, = \, 2 \, z_6 - 2 \, z_3 - 4 \, x_8 - 4 \, x_{14} \, , \end{cases} \nonumber \\
\text{from the {\bf 2352}} \qquad & \begin{cases} 0 \, = \, x_{11} + x_{13} - 32 \, x_4 \, ,\\
0 \, = \, \tfrac{1}{3} \, x_1 - 3 \, x_5 - x_6 + x_{10} - 5 \, x_{11} - x_{12} - x_{13} +6 \, x_{15} \, , \end{cases} \nonumber \\
\text{from the {\bf 3584}} \qquad & \begin{cases} 0 \, = \, - \tfrac{3}{4} \, z_6 - \tfrac{3}{4} \, x_8 - \tfrac{1}{4} \, x_9 + \tfrac{1}{4} \, x_2 + \tfrac{3}{2} \, z_4 + z_5 - z_1 \, , \\
0 \, = \, - \bar{z}_7 + \tfrac{3}{2} \, \bar{z}_4 + \tfrac{1}{4} \, x_6 + \tfrac{3}{4} \, x_{11} - \tfrac{3}{4} \, z_3 - \tfrac{3}{4} \, z_6  \, . \end{cases} \nonumber
\end{align}
We can add to to this set two more equations coming from the equations of motion (\ref{Field equations}) which live in the ${\bf 70}_{+}$. They reads
 \begin{align} \label{Quartic EOM}
 0 & = \tfrac{1}{2} \, x_9 - \tfrac{3}{2} \, x_{14} + (z_1 + \bar{z}_{1}) + \tfrac{3}{4} \, (z_3 + \bar{z}_3 ) + \tfrac{3}{4} \, ( z_4 + \bar{z}_4 ) \, , \notag \\
0 & = \tfrac{3}{4} \, (z_3 + \bar{z}_3 ) + \tfrac{3}{4} \, ( z_4 + \bar{z}_4 ) + \tfrac{9}{16} \, (z_2 + \bar{z}_2 ) + \tfrac{9}{8} \, x_{15} \, . 
\end{align}
The derivation of all these conditions can be found in the appendix.

Using these quartic relations the sGoldstino mass can be simplified to 
\begin{align} \label{sGoldstino mass simplified}
M_{\rm sG}^{2} =  \tfrac{3}{64} \, x_1 + \tfrac{33}{2} \, x_4  - \tfrac{9}{16} \, x_{10} \, .
\end{align}
We will first calculate the sGoldstino mass in a number of known examples of critical points, and subsequently use the quadratic constraints to derive upper and lower bounds on the above expression.

To finish this section, we adress the projection of the mass matrix with the antisymmetric sGoldstino directions. This corresponds to tracing $m^2$ with the projector $\bbQ$, and yields
 \begin{align}
   \bbQ_{ijkl}{}^{mnpq} \, m^2_{mnpq}{}^{ijkl} = \tfrac{1}{12} \, x_1 - \tfrac{1}{4} \, x_2 - \tfrac{3}{8} \, x_5 - \tfrac{1}{16} \, x_6 + \tfrac{1}{48} \, x_7 + \tfrac{9}{16} \, x_{11} - \tfrac{3}{16} \, x_{13} 
 \end{align}
It can be seen that this expression vanishes modulo the quartic relations above. This supports our interpretation of the antisymmetric sGoldstino directions as pure gauge.

\section{Specific examples}

Before going through the discussion about the metastability of critical points in the case of supersymmetry breaking, it is worthwhile to work out the explicit  form of the embedding tensor for some specific gaugings. Some of the easiest examples are built by requiring invariance under maximal subgroups of $SO(8)$, namely $SO(4) \otimes SO(4)$ and $SU(2) \otimes Usp(4)$. Finally we will study four examples that arise in the study of geometric flux compactifications. In what follows we will always choose the normalisation of the embedding tensor such that $|A_1|^{2} = \tfrac{4}{3}$. In this way, the scalar potential in the fully supersymmetric vacuum will be given by $V = -1$.
 \begin{itemize}
 \item {\bf The $SO(8)$ and $SO(4) \otimes SO(4)$ gauging} \\
We split the $i$ index in two $SO(4)$ indices $a$ and $\hat{a}$. The flux components which can be constructed using singlets with respect to these groups are the following
\begin{align*}
A^{ij} & \quad \longrightarrow \quad \begin{cases} A^{ab} = \lambda_1 \, \delta^{ab} \\ \\ A^{\hat{a} \hat{b}} = \hat{\lambda}_1 \, \delta^{\hat{a} \hat{b}} \end{cases} \quad , \qquad 
A_{i}{}^{jkl}  \quad \longrightarrow \quad \begin{cases} A_{a}{}^{bcd} = \lambda_2 \, \delta_{ae} \, \epsilon^{ebcd} \\ \\ A_{\hat{a}}{}^{\hat{b} \hat{c} \hat{d}} = \hat{\lambda}_{2} \, \delta_{\hat{a} \hat{e}} \, \epsilon^{\hat{e} \hat{b} \hat{c} \hat{d}} \end{cases} \, .
\end{align*}
From the quadratic constraints in the $\bf{63}$ we get that 
\begin{align}
|\lambda_1| = | \hat{\lambda}_{1} | \quad , \qquad |\lambda_2| = | \hat{\lambda}_{2} | \, .
\end{align}
Furthermore, from the quadratic constraints in the $\bf{70}$ and the equations of motion, we get two possible solutions, namely either $\lambda_2=0$ or $\lambda_1=0$. 

The first case corresponds to only $A_1$ and therefore the maximally supersymmetric $SO(8)$ gauging that we discussed before. In order to get the correct normalisation $|A_1|^2 = \tfrac43$ and hence $V=-1$ one needs to choose $\lambda_1 = 1/\sqrt{6}$. All scalar masses at this critical point are given by $m^2 = - \tfrac23$.

In the other case the only non zero flux components are $A_{a}{}^{bcd}$ and $A_{\hat{a}}{}^{\hat{b} \hat{c} \hat{d}}$. It necessarily corresponds to the $SO(4,4)$ gauging \cite{HW84}, as this is the unique other gauging that overlaps with the $SO(8)$ gauging in two $SO(4)$ subgroups. For this gauging, being $|A_1|^{2} = 0$, our normalisation fails. Nevertheless, the potential is positive and is given by $V = 2 \, |\lambda_{2}|^{2}$. The mass spectrum is given by
 \begin{align}
 \tfrac{m^2}{V} \, (\text{multiplicity}) : \qquad  -2 \, ( \times 2) \, , \; 0 \, ( \times 16 ) \, , \; 1 \, ( \times 16 ) \, , \; 2 \, ( \times 36) \, .
 \end{align}
 There are two tachyons in the spectrum. They render this de Sitter critical point unstable. Interestingly, the sGoldstino mass is zero in this case.
 
 \item
{\bf The $SU(2) \otimes Usp(4)$ gauging} \\
Another maximal subgroup of $SO(8)$ is  $SU(2) \times Usp(4)$, which is isomorphic to $SO(3) \times SO(5)$. This is the preserved part of the $SO(5,3)$ gauging at the origin. In this case we split the $i$ index in the pair $\alpha M$ with $\alpha = 1,2$ and $M = 1 , \ldots 4$. The flux components are
\begin{align*}
A^{ij} & \quad \longrightarrow \quad A^{\alpha M ,\, \beta N} = \lambda_1 \, \epsilon^{\alpha \beta} \, \Omega^{MN} \, , \\ 
A_{i}{}^{jkl} & \quad \longrightarrow \quad  A_{\alpha M}{}^{\beta N ,\, \gamma P ,\, \delta Q} = \lambda_2 \left( \delta_{\alpha}^{(\beta} \, \epsilon^{\gamma) \delta} \, \delta_{M}^{Q} \, \Omega^{NP} + \delta_{\alpha}^{(\delta} \, \epsilon^{\beta) \gamma} \, \delta_{M}^{P} \, \Omega^{QN} + \delta_{\alpha}^{(\gamma} \, \epsilon^{\delta) \beta} \, \delta_{M}^{N} \, \Omega^{PQ} \right) .
\end{align*}
We find that the quadratic constraints and the equations of motion are satisfied by these decomposition provided
\begin{align*}
|\lambda_2|^{2} = 4 \, |\lambda_1|^{2} \, .
\end{align*}
Our choice for the normalisation fixes $|\lambda_1|^{2} = \tfrac{1}{6}$ and thus $|\lambda_2|^{2} = \tfrac{2}{3}$. The potential is given by $V = 4$. The origin is a de Sitter stationary point with both $A_1$ and $A_2$ turned on. The mass spectrum is given by
 \begin{align}
 \tfrac{m^2}{V} \, (\text{multiplicity}) : \qquad   -2 \, ( \times 1) \, , \; - \tfrac{2}{3} \, ( \times 5) \, , \; 0 \, ( \times 15 ) \, , \; 2 \, ( \times 30 ) \, , \; \tfrac{4}{3} \, ( \times 14 ) \, , \; 4 \, ( \times 5 ) \, .
 \end{align}
Again the presence of a tachyon renders the critical point unstable. The sGoldstino mass is zero also in this case.
\item
{\bf Geometric IIA compactifications} \\
There is a class of half-maximal gauged supergravity which arises as the low energy limit of certain type IIA orientifold compactifications including background fluxes, D6-branes and O6-planes. For this class of theories the complete mass spectrum was worked out in \cite{DGR}. In the same paper it was discovered that the complete vacuum structure can be embedded in the $\cN = 8$ theory. 

The set of vacua is given by 4 points with an additional four-fold degeneracy related to a $\bbZ_2 \times \bbZ_2$ symmetry. We give here the value of the potential, the average mass squared, the most tachyonic field, the stability properties and the amount of residual supersymmetry for these inequivalent four points\footnote{We are grateful to Giuseppe Dibitetto and Adolfo Guarino for providing as of yet unpublished results on these points.} \cite{DGR, DGR2}.
\begin{center}
\begin{tabular}{||c||c|c|c|c|c|c||}
\hline \hline Critical point & $ \quad V \quad $ & Tr$\{m^{2}\}/|V|$ & $m_{\rm sG}^{2}/|V|$ & ${\rm Min}\{m^2 \}/|V|$ & Stable  & SUSY \\
\hline \hline 1 & $-\frac{4}{27}$ & $\frac{12}{5}$ & $ \sim 1.72 $ & $- \frac{2}{3}$ & $\surd$ & $\cN = 1$ \\
\hline 2 & $-\frac{20}{129}$ & $\frac{56}{25}$ & $ \sim 1.54$ & $-\frac{4}{5}$ & $\times$ & -- \\
\hline 3 & $-\frac{4}{33}$ & $\frac{16}{5}$ & $ \sim 2.54 $ & $0$ & $\surd$ & -- \\
\hline 4 & $-\frac{4}{21}$ & $\frac{8}{5}$ & $ \sim 1.12 $ & $-\frac{4}{3}$ & $\times$ & -- \\
\hline \hline
\end{tabular}
\end{center}

\end{itemize}

\section{Bounds on sGoldstino mass}

\subsection{Derivation of bounds}

Using the quartic relations it is possible to simplify the expression (\ref{sGoldstino mass}) for the sGoldstino mass obtaining \eqref{sGoldstino mass simplified}. Despite its simplicity, this expression doesn't have a definite sign. The actual value depends on the interplay between the different terms once a gauging is chosen.
Finding an upper (lower) bound on the sGoldstino mass amounts in maximising (minimising) a function of the independent variables in (\ref{sGoldstino mass simplified}). This function is linear and, as long as the variables are free to move in the whole space, it has no maximum or minimum. Fortunately there are several constraints which can be imposed on those variables by general arguments.

The first, trivial one, comes from the very definition \eqref{xz-space}. We indeed see that 
\begin{align} \label{First inequalities}
x_1,\, x_2 ,\, x_3 ,\, x_5 ,\, x_7 ,\, x_{10} \geq 0
\end{align} 

Another type of constraints comes out considering particular quadratic combinations of the embedding tensor components belonging to irreducible representations. As an example take the following combination
\begin{align*}
A_{r}{}^{smn} A^{r}{}_{sij} - \tfrac{2}{3} \, \delta_{[i}^{[m} \, A_{r}{}^{n]st} A^{r}{}_{j]st} +
\tfrac{1}{21} \, \delta_{ij}^{mn} \, |A_2|^2 \, ,
\end{align*}
belonging to the ${\bf 720}$ irrep. If we take the product of this expression with its complex conjugate, we get by definition a non-negative expression, quartic in $A_{1,2}$. After replacing the dependent variables we are left with
\begin{align*}
A_{r}{}^{smn} A^{r}{}_{sij} \, A_{t}{}^{uij} A^{t}{}_{umn} - \tfrac{8}{3} \left( A_{ir} A^{mr} \, A_{ms} A^{is} - \tfrac{1}{8} |A_{1}|^{2} \, |A_{1}|^{2}  \right) - \tfrac{1}{28} \, \left( |A_2|^{2} \right)^{2} & \geq 0 \\
\longrightarrow \qquad x_{10} - \tfrac{8}{3} \, x_4 - \tfrac{1}{28} \, x_1 & \geq 0 \, .
\end{align*}
In the following the complete list of inequalities obtained using this procedure. 
The actual derivation of all the constraints is shown in the appendices. 

For the $\bf 63$ we find a one-parameter family of inequalities
 \begin{align} \label{Irrep 63}
 | \alpha_{1} |^2 \, x_4 \geq 0 \, .
 \end{align}
For the ${\bf 70}$ we find a two-parameter family
 \begin{align} \label{Irrep 70}
 | \beta_1 |^{2} \, x_{15} - \tfrac{3 \, |\beta_{2} |^{2}}{4} \, x_{14} + \tfrac{|\beta_{2}|^{2}}{4} \, x_9 + \beta_{1} \beta_{2}^{*} \, z_3 + \beta_{1}^{*} \beta_{2} \, \bar{z}_3 \geq 0 \, . 
 \end{align}
The presence of the two parameters in this case is related to the possibility of building different kinds of contractions, using $A_1$ and $A_2$, all belonging to the same irrep. Once again we refer to the appendices for a detailed explaination. For the $\bf 330$ we find a constraint which is already implied by others.
For the $\bf 336$ we find a three-parameter family but the irrep always contains new variables, and hence we will not consider it. For the $\bf 378$ we find a one-parameter family
 \begin{align} \label{Irrep 378}
 |\delta_{1}|^{2} \, ( x_9 + x_{14} ) \geq 0 \, . 
 \end{align}
For the $\bf 720$ we find a three-parameter family
 \begin{align} \label{Irrep720}
 ( |\varepsilon_{1}|^2 + |\varepsilon_{2}|^2 ) \, x_{10} + 2 \, ( \varepsilon_{1} \varepsilon_{2}^{*} + \varepsilon_{1}^{*} \varepsilon_{2} ) \, x_{11} + |\varepsilon_{2}|^{2} \, x_{12} - 2 \, |\varepsilon_{2}|^{2} \, x_{13} + & \\
 - \tfrac{1}{6} \, | -2 \, \varepsilon_{1} + \varepsilon_{2} |^{2} \, x_{5} - \tfrac{1}{6} \, ( - 2 \, \varepsilon_{1}^{*} + \varepsilon_{2}^{*} ) \varepsilon_{2} \, x_{6} - \tfrac{1}{6} \, ( - 2 \, \varepsilon_{1} + \varepsilon_{2} ) \varepsilon_{2}^{*} \, x_6 - \tfrac{1}{6} \, | \varepsilon_{2} |^{2} \, x_7 + & \nonumber \\
 + \tfrac{| \varepsilon_{1} -  \varepsilon_{2}|^{2}}{21} \, x_1 + ( \varepsilon_{1} +  \varepsilon_{2} ) \, \varepsilon_{3}^{*} \, \bar{z}_{6} + ( \varepsilon_{1}^{*} +  \varepsilon_{2}^{*} ) \, \varepsilon_{3} \, z_{6} + \varepsilon_{2} \varepsilon_{3}^{*} \, \bar{z}_3 + \varepsilon_{3} \varepsilon_{2}^{*} \, z_3 + \tfrac{|\varepsilon_{3}|^{2}}{2} \, x_8 - \tfrac{|\varepsilon_{3}|^{2}}{2} \, x_{14} & \geq 0 \, . \nonumber 
 \end{align}
For the $\bf 945$ we find a one-parameter family
\begin{align} \label{Irrep 945}
 | \zeta_1 |^{2} ( x_8 + x_{14} ) \geq 0 \, .
\end{align}
For the $\bf 1232$ we find a two-parameter family
 \begin{align} \label{Irrep 1232}
 |\eta_{1}|^{2} \, x_{10} + |\eta_{1}|^{2} \, x_{12} + 2 \, |\eta_{1}|^{2} \, x_{13} - \tfrac{|\eta_{1}|^{2}}{10} \, ( x_{5} + 2 \, x_6 + x_7 ) + & \\
 - \tfrac{\eta_{1} \eta_{2}^{*} + \eta_{2} \eta_{1}^{*}}{5} \, ( x_{8} + x_{9} ) + 2 \, (\eta_{1} \eta_{2}^{*} + \eta_{2} \eta_{1}^{*}) \, x_{14} + & \nonumber \\
 \tfrac{|\eta_{1}|^{2}}{45} \, x_1 + \tfrac{\eta_{1} \eta_{2}^{*} + \eta_{2} \eta_{1}^{*}}{45} \, x_2 + \tfrac{35 \, |\eta_{2}|^{2}}{36} \, x_3 - \tfrac{2 \, |\eta_{2}|^{2}}{5} \, x_4 & \geq 0 \, . \nonumber
 \end{align}
For the $\bf 1764$ we find a one-parameter family
\begin{align} \label{Irrep 1764}
 |\theta_1|^{2} \, \left( x_{10} - 5 \, x_{11} + \tfrac{1}{2} \, x_{12} - \tfrac{5}{2} \, x_{13} - \tfrac{9}{8} \, x_5 - \tfrac{10}{8} \, x_6 - \tfrac{1}{8} \, x_7 + \tfrac{1}{5} \, x_1 + \tfrac{9}{2} \, x_{15} \right) \geq 0 \, .
\end{align}
For the $\bf 2352$ we find a one-parameter family
 \begin{align} \label{Irrep 2352}
 |\iota_1|^{2} \, \left( - \tfrac{9}{4} \, x_{10} + \tfrac{9}{10} \, x_5 + x_7
- \tfrac{1}{20} \, x_1 \right) \geq 0 \, . 
 \end{align}
For the $\bf 3584$ we find a one-parameter family
 \begin{align} \label{Irrep 3584}
 |\kappa_{1}|^{2} (  x_6 + 3 \, x_{11} - 2 \, x_{15} ) \geq 0 \, .
 \end{align}

Moreover, one can employ a number of ``matrix tricks'' to derive further bounds on the domain of the sGoldstini mass. Whenever we have a matrix
\begin{align*}
M_{I}{}^{J} = V_{I} \, V^{J}  \,, 
\end{align*}
with $V^{J} = (V_{J})^{*}$,  it is hermitean and has real, non-negative eigenvalues. Hence one can write the following inequality
\begin{align*}
\Tr\{M^2 \} \leq [\Tr\{M\}]^{2}  \,.
\end{align*}
The left hand side is nothing but $\sum_{I=1}^{n} \lambda_{I}^{2}$ while the right hand side is $(\sum_{I=1}^{n} \lambda_{I})^{2}$ with $\lambda_{I}$ being the non negative eigenvalues. Applying this to $A_{ir} A^{mr}$, $A_{r}{}^{smn} A^{r}{}_{sij}$, $A_{r}{}^{stm} A^{r}{}_{sti}$ and $A_{i}{}^{rst} A^{m}{}_{rst}$ one finds respectively
 \begin{align} \label{Matrix tricks xz}
  x_1 - x_{10} \geq 0 \, , \quad \tfrac{7}{8} \, x_3 - x_4 \geq 0 \, , \quad x_1 - x_5 \geq 0 \, , \quad x_1 - x_7 \geq 0 \, .
 \end{align}

The last type of constraints we have been able to find has a geometric origin. In particular, the $E_7/SU(8)$ scalar manifold is a symmetric space with the interesting property that, at any point, the sectional curvature is always non positive\cite{Helgason}.
The sectional curvature is defined taking a suitable contraction of the Riemann tensor with vectors, spanning a plane in the tangent space. We have two sets of directions along the scalar manifold, namely the symmetric and antisymmetric sGoldstino directions. From those directions we can construct the following three quantities proportional to sums of sectional curvatures
\begin{align} \label{Sectional curvature xz}
R \, \bbQ \, \bbQ & = -6 \, x_1 + 9 \, x_5 - 15 \, x_6 - 3 \, x_7 - 27 \, x_{10} + 18 \, x_{11} + 45 \, x_{13} + 27 \, x_{15} \, , \nonumber \\
R \, \bbQ \, \bbP & = -10 \, x_1 + 45 \, x_5 + 3 \, x_6 - x_7 - 27 \, x_{10} + 18 \, x_{11} + 9 \, x_{13} - 27 \, x_{15} \, , \nonumber \\
R \, \bbP \, \bbP & = -18 \, x_1 + 81 \, x_5 + 9 \, x_6 + 5 \, x_7 - 27 \, x_{10} - 54 \, x_{11} - 63 \, x_{13} + 27 \, x_{15} \, .
\end{align}
As explained before, these quantities are non positive due to the geometry of the scalar coset space.

This finishes our derivation of the quartic inequalities that we will use. Note that this is not necessarily an euxhaustive list but we will restrict ourselves to the above in what follows. In particular, We will now turn to maximising (minimising) the function (\ref{sGoldstino mass simplified}) subjected to (\ref{Irrep 63}) up to \eqref{Sectional curvature xz}. Geometrically these constraints define a domain in the variable space. The problem is a generalisation of a constrained extremalisation problem where the constraints are now inequalities instead of equalities.

\subsection{The case $A_1 = 0$}
For all the gaugings with $A_1=0$, in a critical point, we have a positive cosmological constant, thus a de Sitter space. All the known examples of de Sitter configurations in $\cN = 8$ supergravity are unstable. We will see that, the sGoldstino direction in this case inherits some of the instability but it's not parallel to the tachyonic directions. Indeed it turns out that its value will always be 0.

The proof is simple and goes as follows. The sGoldstino mass (\ref{sGoldstino mass simplified}) reduces to
\begin{align} \label{sGoldstino mass simplified A1=0}
M_{\rm sG}^{2} =  \tfrac{3}{64} \, x_1  - \tfrac{9}{16} \, x_{10} \, ,
\end{align}
thus its sign is still uncertain due to the interplay between a negative and a positive contribution. Nevertheless we can see that, taking (\ref{Irrep720}), putting to zero all the terms proportional to $A_1$ and choosing $\varepsilon_1=\varepsilon_2=1$, we get
\begin{align*}
- \tfrac{1}{4} \, x_1 + 3 \, x_{10} \geq 0 \, .
\end{align*}
On the other hand, in the absence of $A_1$, in terms of the independent variables the inequality (\ref{Irrep 1764}) reads
\begin{align*}
\tfrac{3}{16} \, x_1 - \tfrac{9}{4} \, x_{10} \geq 0 \, .
\end{align*}
Both the above inequalities have to hold. The first one sets $0$ as an upper bound for the sGoldstino mass, while the second one sets $0$ as a lower bound. 

This tells us that the sGoldstino mass necessarily vanishes. In other words, the allowed window between the minimum and the maximum of the sGoldstino mass shrinks to a point in the case where $A_1$ vanishes. This coincides with our previous result for the $SO(4,4)$ gauging, but we now find that this holds for all critical points where $A_1 = 0$. Therefore, in the case where the scalar potential is completely determined by the order parameter of supersymmetry breaking, the mass spectrum will always contain either a number of tachyonic modes, or several flat directions. A stable De Sitter vacuum with strictly positive masses is impossible.

\subsection{The general case}

In the general case the analysis is somewhat more complicated. The presence of $A_1$ opens up more directions in the $\vec{x}$ space. It's unclear which of the inequalities amongst (\ref{Irrep 63})-(\ref{Sectional curvature xz}) will give the most stringent bound on the variables appearing in the sGoldstino mass. In other words, the domain defined by all the constraints has a non-trivial shape due to the large number of variables. The problem is really a generalisation of the constrained maximisation (minimisation) of a function. Fortunately there are some algorithms which allow one to solve this kind of problems. We have employed the one built in in Mathematica. In this way we have calculated  the maximum and the minimum of the sGoldstino mass \eqref{sGoldstino mass simplified}, whenever the independent $\vec{x}$ and $\vec{z}$ variables are subject to all the constraints given above.

The only technical point resides in the inequalities which depend on two or more parameters and must hold for whichever value of these parameters (see for instance (\ref{Irrep 70}), (\ref{Irrep720}) etc.). If we use Mathematica to maximise (or minimise) the sGoldstino mass subjected to such a parametric constraint, the computing time blows up irremediably. To solve this issue we've chosen to generate from those general inequalities containing two parameters, a number of resulting inequalities by fixing the value of the parameters. This constitutes of course a loss of generality but it's worth to mention two points. 

First of all, from a conceptual point of view, this procedure amounts to soften the constraints on the allowed domain. In other words there can be more space for the independent variables and thus the maximum (minimum) of the function can only be higher (lower). If we find an interesting bound in a larger domain things could only improve if we were able to account for the full, parametric constraint. Nevertheless we've tried to generate a huge number of resulting inequalities in order to constrain as much as possible the problem. Furthermore, we have noticed that, for different numbers and choices of resulting inequalities, the value of the maximum and the minimum does not change as long as one includes a minimum amount of constraints.

 \begin{figure}[h!]
\centering
\includegraphics[angle=0,width=0.77\textwidth]{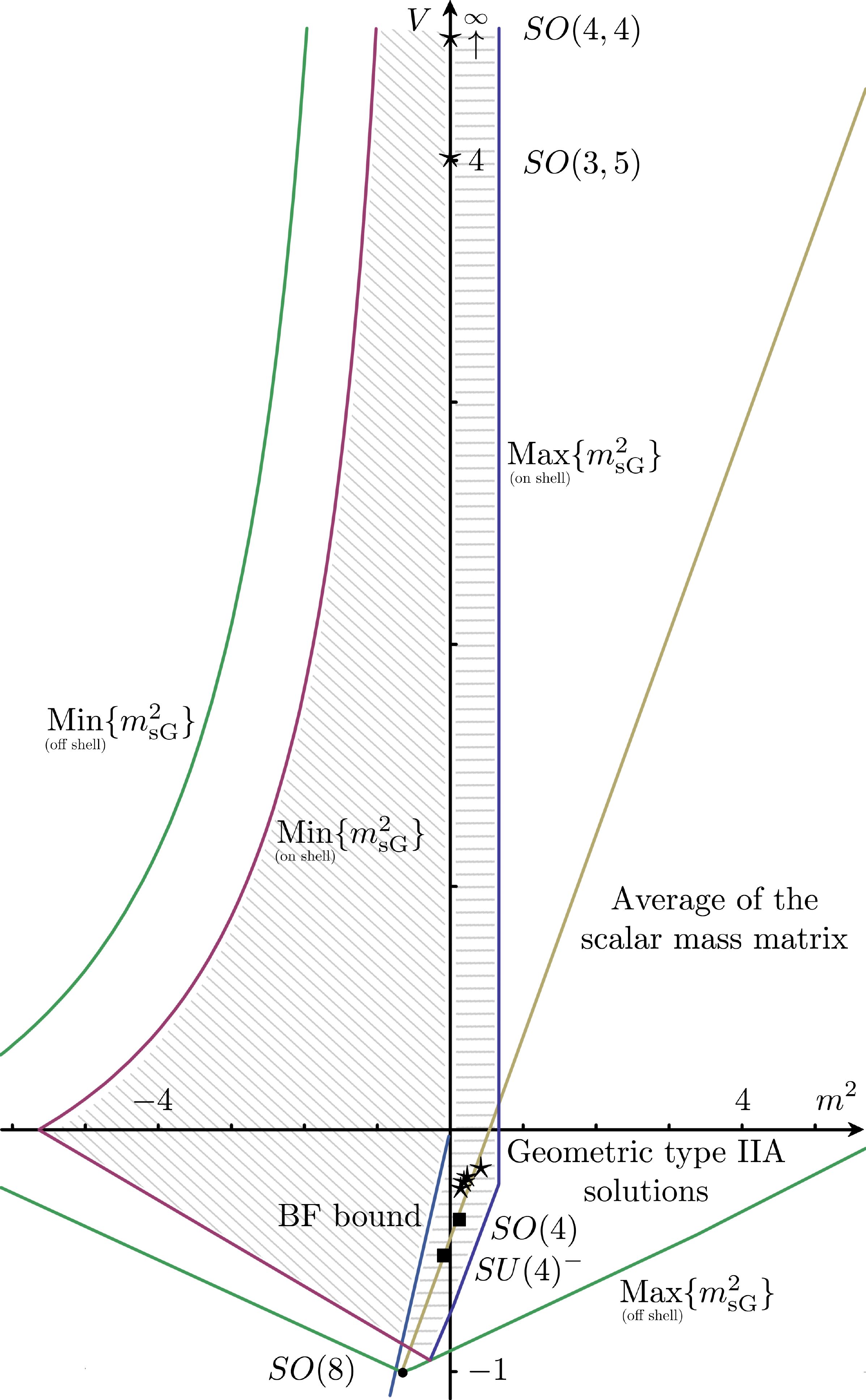}
\caption{The bounds on the sGoldstino mass (horizontally) in units of $\tfrac34 |A_1{}|^2$ as a function of the scalar potential (vertically). The different symbols are explained on the previous page.}
\end{figure}

We are thus finally able to give the results. They are summarised in figure 1, in which the different symbols signify the following:
 \begin{itemize}
 \item
  On the $x$-axis is the mass of scalar fields, normalised with respect to $\tfrac{3}{4} \, |A_1|^2$. On the $y$-axis we have plotted the value of the scalar potential in the normalisation where $|A_1|^{2} = \tfrac{4}{3}$, which is chosen in order to have $V = -1$ in the fully supersymmetric case. 
 \item
 We have  included the straight line corresponding to the average over all 70 scalars, whose mass is given by \eqref{Average mass}. This corresponds to the normalised trace over the mass matrix.
 \item
  In the Anti-de Sitter part, we have also depicted the the BF bound which divide this region in a stable and an unstable sector.
 \item
 The allowed window for the sGoldstino mass is delimited by the two lines Min$\{m_{\rm sG}^{2}\}$ and Max$\{m_{\rm sG}^{2}\}$. In fact, we have included two such windows: an off-shell window, where we only incorporate the consequences of the quadratic constraints, and an on-shell window, where we additionally include the quartic relations following from the field equations in the ${\bf 70}_+$. Clearly the on-shell window sits inside the off-shell one. Furthermore, we have indicated the stable sGoldstino by a filling with horizontal lines, while the unstable region is filled with diagonal lines.
 \item
 Finally, we have included the sGoldstino masses of the examples discussed in section 4 (i.e.~$SO(4,5)$, $SO(3,5)$ and geometric IIA) with a star symbol. In addition, we have included the $SU(4)^{-}$ and $SO(4)$ critical points of \cite{Warnernew1, Warnernew2} by putting a black box for their averaged masses. The sGoldstino masses are not known in these cases.
\end{itemize}
A number of interesting conclusions can be drawn from this picture.

Let's start with the upper part of the picture ($V > 0$). We see that the maximum stays constant while the minimum goes to 0. This is an interesting finding. Indeed if one had computed $m_{\rm sG}^{2}/V$ instead of normalising $m_{\rm sG}^{2}$ to $\tfrac{3}{4} \, |A_1|^2$, the window would have shrunk to zero. In other words, the further one moves in the de Sitter region ($|A_2|^2 \gg |A_1|^2$) the more $\eta_{\rm sG} \equiv m^{2}_{\rm sG}/V$ approaches zero. Remember that the sGoldstino mass is roughly an average over several masses and the fact that it's almost zero implies either the presence of a number of light modes or the presence of tachyons. First of all, this exemplefies the difficulty in finding stable De Sitter. In particular, it points out the lower region of the upper half plane of this picture as most favorable in this respect. Secondly, this demonstrates the power of the method based on the sGoldstino mass, especially when compared to the average of the scalar mass matrix. The latter quantity diverges as one increases $|A_2|^2$, while the maximum for the sGoldstino stays constant. 

 The lower part of the picture ($V < 0$) is equally interesting. A large part of the allowed region sits in the unstable sector while there is a small strip between the BF bound and the maximum of the sGoldstino mass. Inside this region we see the four sGoldstino masses corresponding to the geometric type IIA solutions of section 4. Unfortunately here we see a drawback of our method. Indeed, while not all these solutions are stable, the sGoldstino masses always sit in the stable sector of the window. This explains our statement that the sGoldstino captures part of the instability but it's not parallel to the most unstable direction in the scalar coset. 

There is a further important piece of information which can be read from the picture. Indeed we see that the on-shell window closes up around $(m^2, \, V) = (- 0.27, \, -0.93)$, which is a finite distance above the fully supersymmetric critical point indicated by a bullet at $(m^2, \, V) = (-2/3, -1)$. This means there is a gap between the $\cN = 8$ supersymmetric critical point and whichever other critical point in which supersymmetry is partially or completely broken. In other words, in maximal gauged supergravity, it is impossible to go smoothly from a maximally supersymmetric critical point to a non supersymmetric one along a critical path. As mentioned before, this is a consequence of the quadratic constraints and the field equations: there are no solutions to this combined system which allow for a smooth limit to the maximally supersymmetric solution\footnote{The same conclusion was reached for $\cN = 4$ supergravity via a somewhat different line of reasoning related to SUSY mass terms \cite{BR}.}. 

From the generality of the argument based on the quadratic constraints, one may anticipate that this holds true for all supersymmetric critical points: none of these allow for a smooth deformation into non-supersymmetric critical points. This would imply a minimal amount of supersymmetry breaking, and hence the absence of the notion of approximate supersymmetry, in maximal supergravity.

\section{Conclusions}

This paper addresses a necessary condition for stability of all the critical points of maximal supergravity apart from the $SO(8)$ theory in the origin, preserving maximal supersymmetry. All other critical points break either a fraction or all of supersymmetry, and in the latter cases it has proven difficult to realise stability. Our findings explain why this has been such a hard task: of the full 912-dimensional parameter space, only a small fraction allows for a sGoldstino mass that is positive or above the BF bound. Furthermore the sGoldstino only furnishes a necessary condition, and hence stability with respect to all 70 scalars is yet harder to realise. It should also be borne in mind that our results, and in particular the window of figure 1, also includes partially supersymmetric critical points in the AdS region.

Possible extensions of this research include additional constraints on the embedding tensor, narrowing the search. For instance, if one would only be interested in the $SO(8)$ theory and its properties, one could look for additional quadratic relations that the embedding tensor everywhere in the moduli space of the $SO(8)$ theory satisfies. However, one should not expect these relations to be $SU(8)$ covariant for the following reason. Suppose they can be formulated in $SU(8)$ language, i.e.~are covariant with respect to the compact generators of $E_{7(7)}$. The requirement that they are true in all of moduli space imposes covariance with respect to the non-compact generators; in other words, they would have to form $E_{7(7)}$ irreps. At the quadratic level in the embedding tensor, there are five such irreps:
 \begin{align}
  ({\bf 912} \times {\bf 912})_{\rm symm} = {\bf 133} + {\bf 8645} + {\bf 1463} + {\bf 152152} + {\bf 253935} \,.
 \end{align}
The first two of these correspond to quadratic constraints, satisfied by all theories. We have explicitly checked that the other three irreps do not vanish for either the $SO(8)$, $SO(4,4)$, $SO(5,3)$ and geometric IIA examples. Therefore one would have to go to e.g.~quartic level to find $SU(8)$ covariant expressions that restrict one to a specific theory such as the $SO(8)$ one. There is a singlet in the four-tuple symmetric product of the ${\bf 912}$. Imposing this to vanish would correspond to an additional hypersurface in the space of quartic variables, and thus narrow down the range of the sGoldstino mass.

Other interesting future research directions include a refinement of our analysis by also considering scalar partner of the would-be Goldstone boson in the case of gauge symmetry breaking, in addition to the Goldstino partner in the case of SUSY breaking \cite{Brizi:2011jj}. These are clearly intertwined in the present case, as only the $SO(8)$ critical point preserves both all supersymmetry and gauge symmetry. Similarly, it would be interesting to analyse inflationary properties of the scalar potential of maximal supergravity along the lines of \cite{Covi2}. Finally, our method of analysing the constraints on quartic relations of the gauge parameters and the resulting consequences for the sGoldstino mass can be applied to other cases, including e.g. the open problem of $\cN = 2$ supergravity coupled to vector multiplets.

\section*{Acknowledgements}

We would like to thank Giuseppe Dibitetto, Adolfo Guarino, Diego Marques, Jan Rosseel and Claudio Scrucca for very useful and interesting discussions. The work of A.B.~and D.R.~is supported by a VIDI grant from the Netherlands Organisation for Scientific Research (NWO).  R.~L.~acknowledges support from RED-FAE CONACyT and the University of Groningen for warm hospitality.

\appendix
\section{Derivation of the constraints on the coordinates}

\subsection{Quadratic constraints}

Starting from the quadratic constraints in every irreducible representation and taking the product with a complex conjugate representation we can construct a series of equalities which are quartic in the embedding tensor components. In order to do so in an exhaustive way, one needs to look at the products of two embedding tensors, and see which of the resulting irreps are complex conjugate to a quadratic constraint. The list of tensor products reads
 \begin{align}
   ({\bf 36} \times {\bf 36})_{\rm s} & = {\bf 330} + {\bf 336} \,, \notag \\
  {\bf 36} \times \overline {\bf 36} & = {\bf 1} + {\bf 63} + {\bf 1232} \,, \notag \\
   ({\bf 420} \times {\bf 420})_{\rm s} & = {\bf 70} + {\bf 336} + \overline{\bf 336} + \overline{\bf 378} + {\bf 3584} +  \ldots \,, \notag \\
  {\bf 420} \times \overline {\bf 420} & = {\bf 1} + 2 \cdot {\bf 63} + 2 \cdot {\bf 720} + {\bf 945} + \overline{\bf 945} + {\bf 1232} + {\bf 1764} + 2 \cdot {\bf 2352} + \ldots  \,, \notag \\
  {\bf 36} \times {\bf 420} & = {\bf 70} + {\bf 378} + {\bf 3584} + \ldots \,, \notag \\
  {\bf 36} \times \overline {\bf 420} & = {\bf 720} + {\bf 945} + \ldots \,,  \label{product}
  \end{align}
where we have only explicitly given the irreps up to and including dimension $3584$. Then we could take the following products
\begin{align} \label{obtaining quartic relations}
{\bf 63} \quad & \times \quad A_{u}{}^{vzi} A^{u}{}_{vzm} \, , \quad A_{m}{}^{uvz} A^{i}{}_{uvz} \, , \quad A_{mu} A^{iu} \, , \nonumber \\
{\bf 70}_- \quad & \times \quad A_{t}{}^{u[ij} A_{u}{}^{t|kl]} \, , \quad A_{t}{}^{[ijk} A^{l]t} \, , \nonumber \\
{\bf 378} \quad & \times \quad \epsilon^{abcdeijk} A^{l}{}_{def} A^{f}{}_{abc} \, , \quad A_{t}{}^{ijk} A^{lt} \, , \nonumber \\
{\bf 945} \quad & \times \quad A_{m}{}^{tui} A^{j}{}_{ntu} \, , \quad A_{mn} A^{ij} \, , \nonumber \\
{\bf 2352} \quad & \times \quad A_{t}{}^{ijk} A^{t}{}_{mnp} \, , \quad A_{[m}{}^{t[ij} A^{k]}{}_{np]t} \, , \nonumber \\
{\bf 3584} \quad & \times \quad A_{i}{}^{ja[m} A_{a}{}^{npq]} \, , \quad A_{i}{}^{[mnp} A^{q]j}
\end{align}
In the main body of the paper we have reported in (\ref{Quartic relations complete}) only the independent relations obtained in this way. Regarding the equations of motion, the procedure is almost the same. Taking the product
 \begin{align} 
 \text{equations of motion} \quad & \times \quad A_{t}{}^{u[ij} A_{u}{}^{t|kl]} \, , \quad A_{t}{}^{[ijk} A^{l]t} \, ,
 \end{align}
 we get the equations (\ref{Quartic EOM}). 
 
 \subsection{Constraints coming from irreps}
 Turning to the irreps, in principle we can use any of the irreps in the tensor product \eqref{product}, take the product with the complex conjugate, and obtain a quadratic expression in $\vec{x}$ and $\vec{z}$ which must be positive. However, we can always leave out the quadratic constraints, as these will only lead to quartic inequalities that are saturated by the quartic relations derived above and listed in (\ref{Quartic relations complete}). 
 
 It's worth to give an example. Take for instance the case of the ${\bf 63}$. We could write a three parameter combination
 \begin{align*}
 \alpha _1 \, A_{i}{}^{rst} A^{m}{}_{rst} + \alpha_2 \, A_{r}{}^{stm} A^{r}{}_{sti} + \alpha_3 \, A_{ir} A^{mr} - \tfrac{1}{8} \, (\alpha_1 + \alpha_2) \, \delta_{i}^{m} \, |A_2|^{2} - \tfrac{1}{8} \, \alpha_3 \, \delta_{i}^{m} \, |A_1|^{2} \, ,
 \end{align*}
 belonging to this irrep. However, we can always use the two independent quadratic constraints in (\ref{Quadratic constraints I}) to express everything in terms of a one parameter family. For instance
 \begin{align}
 \alpha \, A_{ir} A^{mr} - \tfrac{1}{8} \, \alpha \, \delta_{i}^{m} \, |A_1|^{2} \, .
 \end{align}
 Therefore the number of independent combinations (and hence the number of parameter) appearing in any constraint will be obtained taking the non-trivial irreps in \eqref{product} and subtracting the number of quadratic constraints in that irreps. We list here the independent combinations in every irrep. The inequalities (\ref{Irrep 63})-(\ref{Irrep 3584}) are obtained taking the product with the complex conjugate expression.

For the ${\bf 70}$ we find a two-parameter family
 \begin{align}
 \beta_{1} \, A^{r}{}_{s[ij} A^{s}{}_{r|kl]} + \beta_2 \, A^{r}{}_{[ijk} A_{l]r} \, .
 \end{align}
 As already mentioned in the text, the constraints coming from the ${\bf 330}$ are implied by the others while in those coming from the ${\bf 336}$ there are other variables, not present in the list (\ref{xz-space}). Adding them will not give any additional information. Thus we do not consider them. For the ${\bf 378}$ we find a one parameter combination
 \begin{align}
 \delta_1 \, A^{r}{}_{ijk} A_{lr} + \delta_1 \, A^{r}{}_{l[ij} A_{k]r} \, .
 \end{align}
 In the ${\bf 720}$ the situation is a bit more complicated because we don't have any quadratic constraint dwelling in this irrep. Thus we have to consider the full three parameter combination given by
 \begin{align}
& \varepsilon_{1} \, A_{r}{}^{smn} A^{r}{}_{sij} + 2 \, \varepsilon_{2} \, A_{[i}{}^{rs[m} A^{n]}{}_{j]rs} + \tfrac{1}{3} \, (- 2 \, \varepsilon_1 + \varepsilon_2) \, \delta_{[i}^{[m} \, A_{r}{}^{n]st} A^{r}{}_{j]st} + \nonumber \\
& + \tfrac{1}{3} \, \varepsilon_2 \, \delta_{[i}^{[m} \, A_{j]}{}^{rst} A^{n]}{}_{rst} + \tfrac{1}{21} \, (\varepsilon_{1} - \varepsilon_{2}) \, \delta_{ij}^{mn} \, |A_{2}|^{2} + \varepsilon_3 \, A^{[m}{}_{ijr} A^{n]r} \, .
 \end{align}
 Taking the product we find (\ref{Irrep720}). For the ${\bf 945}$ it's sufficient to take the term
 \begin{align}
 \zeta_1 \, A^{(m}{}_{ijr} A^{n)r} \, .
 \end{align}
In the ${\bf 1232}$ again there are no quadratic constraints and we need to consider the following expression
\begin{align}
& 2 \, \eta_{1} \, A_{(i}{}^{rs(m} A^{n)}{}_{j)rs} - \tfrac{1}{5} \, \eta_1 \, \delta_{(i}^{(m} \, A_{r}{}^{n)st} A^{r}{}_{j)st} - \tfrac{1}{5} \, \eta_1 \, \delta_{(i}^{(m} \, A_{j)}{}^{rst} A^{n)}{}_{rst} + \tfrac{1}{45} \, \eta_1 \, \delta_{(i}^{(m} \delta_{j)}^{n)} \, |A_{2}|^{2} + \nonumber \\
& + \eta_2 \, A_{ij} A^{mn} - \tfrac{2}{5} \, \eta_2 \, \delta_{(i}^{(m} \, A_{j)r} A^{n)r} + \tfrac{1}{45} \, \eta_2 \, \delta_{(i}^{(m} \delta_{j)}^{n)} \, |A_{1}|^{2} \, .
\end{align} 
For the $\bf 1764$ we find a one-parameter family
\begin{align}
& \theta_1 \, A_{[i}{}^{[mnp} A^{q]}{}_{jkl]} + \tfrac{1}{2} \, \theta_1 \, \delta_{[i}^{[m} \, A_{r}{}^{npq]} A^{r}{}_{jkl]} + \tfrac{9}{2} \, \theta_1 \, \delta_{[i}^{[m} \, A_{j}{}^{r|np} A^{q]}{}_{kl]r} + \nonumber \\
& - \tfrac{3}{2} \, \theta_{1} \, \delta_{[ij}^{[mn} \, A_{r}{}^{pq]s} A^{r}{}_{kl]s} + 3 \, \theta_1 \,  \delta_{[ij}^{[mn} \, A_{k}{}^{rs|p} A^{q]}{}_{l]rs} + \nonumber \\
& + \tfrac{3}{4} \, \theta_1 \, \delta_{[ijk}^{[mnp} \, A_{r}{}^{q]st} A^{r}{}_{l]st} + \tfrac{1}{4} \, \theta_1 \, \delta_{[ijk}^{[mnp} \, A_{l]}{}^{rst} A^{q]}{}_{rst} - \tfrac{1}{20} \, \theta_1 \, \delta_{ijkl}^{mnpq} \, |A_{2} |^{2} \, .
\end{align}
For the $\bf 2352$ we find a one-parameter family
\begin{align}
\iota_1 \, A_{r}{}^{mnp} A^{r}{}_{ijk} - \tfrac{9}{4} \, \iota_1 \, \delta_{[i}^{[m} \, A_{r}{}^{np]s} A^{r}{}_{jk]s} + \tfrac{9}{10} \, \iota_1 \, \delta_{[ij}^{[mn} \, A_{r}{}^{p]st} A^{r}{}_{k]st} - \tfrac{1}{20} \, \iota_1 \, \delta_{ijk}^{mnp} \, |A_{2}|^{2} \, .
\end{align}
And finally for the $\bf 3584$ we find a one-parameter family
\begin{align}
\kappa_1 \, A^{i}{}_{jv[m} A^{v}{}_{npq]} + \tfrac{1}{12} \, \kappa_1 \, \delta^{i}_{j} \, A^{r}{}_{s[mn} A^{s}{}_{s|pq]} - \tfrac{2}{3} \, \kappa_1 \, \delta^{i}_{[m} \, A^{r}{}_{sj|n} A^{s}{}_{pq]r} \, . 
\end{align}

\subsection{Riemann tensor}
The Riemann tensor for the coset space $E_{7}/SU(8)$ can be written in the form
\begin{align} \label{Riemann tensor in general}
R^{A}{}_{BDE} = \tfrac{1}{4} \, f^{A}{}_{BC} \, f^{C}{}_{DE} + \tfrac{1}{2} \, f^{A}{}_{BI} \, f^{I}{}_{DE} + \tfrac{1}{8} \, f^{A}{}_{CD} \, f^{C}{}_{BE} - \tfrac{1}{8} \, f^{A}{}_{CE} \, f^{C}{}_{BD} \, ,
\end{align}
where $f$ are the structure constants of the group $E_{7}$, the indices $A,B, \ldots$ refer to the non-compact generators while the indices $I, J, \ldots$ refer to the compact ones. \\
Thus, in our case, in $SU(8)$ notation, ${}^{A} \equiv {}^{[ijkl]}$ and ${}^{I} \equiv {}^{i}{}_{j}$. In order to obtain the structure constants we use the definition
\begin{align} \label{Definition structure constants}
[t_{A}] \, f^{A}{}_{BC} = \big[ \, [t_{B}] , \, [t_{C}] \, \big] \, .
\end{align}
The explicit form of the generators in the fundamental representation is the following
\begin{align} \label{Generators E7}
[t_{ijkl}] & = \begin{matr}{cc} 0 & (t_{ijkl})_{mnpq} \\ (t_{ijkl})^{mnpq} & 0 \end{matr} = \begin{matr}{cc} 0 & \tfrac{1}{24} \epsilon_{ijklmnpq} \\ \delta_{ijkl}^{mnpq} & 0 \end{matr} \nonumber \, , \\
[t_{i}{}^{j}] & = \begin{matr}{cc} (t_{i}{}^{j} )_{mn}{}^{pq} & 0 \\ 0 & (t_{i}{}^{j})^{mn}{}_{pq} \end{matr} = \begin{matr}{cc} - \delta_{[m}^{j} \, \delta_{n]i}^{pq} - \tfrac{1}{8} \delta_{i}^{j} \delta_{mn}^{pq} & 0 \\ 0 & \delta_{[p}^{j} \, \delta_{q]i}^{mn} + \tfrac{1}{8} \delta_{i}^{j} \delta^{mn}_{pq} \end{matr} \, .
\end{align}
This gives the following equations for the structure constants
\begin{align} \label{Equations for the structure constants}
[t_{i_1 j_1 k_1 l_1}] \, f^{i_1 j_1 k_1 l_1}{}_{, \,  i_2 j_2 k_2 l_2 , \, i_3 j_3 k_3 l_3} & = \big[ \, [t_{i_2 j_2 k_2 l_2}] , \, [t_{i_3 j_3 k_3 l_3}] \, \big] \nonumber \, , \\
[t_{i_1 j_1 k_1 l_1}] \, f^{i_1 j_1 k_1 l_1}{}_{, \, i_2}{}^{j_2}{}_{, \, i_3 j_3 k_3 l_3} & = \big[ \, [t_{i_2}{}^{j_2}] , \, [t_{i_3 j_3 k_3 l_3}] \, \big] \nonumber \, , \\
[t_{i_1}{}^{j_1}] \, f^{i_1}{}_{ j_1}{}_{, \, i_2 j_2 k_2 l_2 , \, i_3 j_3 k_3 l_3} & = \big[ \, [t_{i_2 j_2 k_2 l_2}] , \, [t_{i_3 j_3 k_3 l_3}] \, \big]  \, .
\end{align}
From the very form of the generator we can argue that
\begin{align} \label{Condition on the structure constants}
f^{i_1 j_1 k_1 l_1}{}_{, \,  i_2 j_2 k_2 l_2 , \, i_3 j_3 k_3 l_3} = 0 \, .
\end{align}
This can be phrased in other words saying that $E_7/SU(8)$ is a symmetric space as the commutator between two non-compact generators is proportional to a compact generator $[ \mathfrak{k} ,\, \mathfrak{k}] = \mathfrak{h}$. Moreover the expression for the Riemann tensor gets a bit simplified.
\begin{align}
R^{i_1 j_1 k_1 l_1}{}_{,\, i_2 j_2 k_2 l_2 , \, i_3 j_3 k_3 l_3 , \, i_4 j_4 k_4 l_4} & = \tfrac{1}{2} \, f^{i_1 j_1 k_1 l_1}{}_{,\, i_2 j_2 k_2 l_2 , \, m_1}{}^{n_1} \, f^{m_1}{}_{n_1 ,\, i_3 j_3 k_3 l_3 , \, i_4 j_4 k_4 l_4} \nonumber \\
& = \; - \tfrac{1}{2} \, f^{i_1 j_1 k_1 l_1}{}_{,\, m_1}{}^{n_1}{}_{, \, i_2 j_2 k_2 l_2} \, f^{m_1}{}_{n_1 ,\, i_3 j_3 k_3 l_3 , \, i_4 j_4 k_4 l_4} \, .
\end{align} 
Applying (\ref{Equations for the structure constants}) and using the explicit form of the Cartan-Killing metric $\kappa_{\alpha \beta} = \mathrm{Tr}\{ t_{\alpha} \, t_{\beta} \}$
\begin{align} \label{Cartan-Killing metric}
\kappa_{ijkl,\,mnpq} = \tfrac{1}{12} \, \epsilon_{ijklmnpq} \qquad , \qquad \qquad \kappa_{i}{}^{j}{}_{,\,k}{}^{l} = 3 \, (\delta_{i}^{l} \delta_{k}^{j} - \tfrac{1}{8} \delta_{i}^{j} \delta_{k}^{l} )  \, ,
\end{align} 
we can extract the remaining structure constants
\begin{align} \label{Structure constants}
f^{i_1 j_1 k_1 l_1}{}_{,\, m_1}{}^{n_1}{}_{, \, i_2 j_2 k_2 l_2} & = 2 \left( \delta_{m_1 [i_2 j_2 k_2}^{i_1 j_1 k_1 l_1} \delta_{l_2]}^{n_1} + \tfrac{1}{8} \delta_{i_2 j_2 k_2 l_2}^{i_1 j_1 k_1 l_1} \delta_{m_1}^{n_1} \right) \nonumber \, ,\\
f^{m_1}{}_{n_1 ,\, i_3 j_3 k_3 l_3 , \, i_4 j_4 k_4 l_4} & = \tfrac{1}{36} \left( \epsilon_{i_4 j_4 k_4 l_4 n_1 [i_3 j_3 k_3} \delta_{l_3]}^{m_1} -  \epsilon_{i_3 j_3 k_3 l_3 n_1 [i_4 j_4 k_4} \delta_{l_4]}^{m_1} \right) \, .
\end{align}
These lead to
\begin{align} 
R^{m_1 n_1 p_1 q_1 ,\, m_2 n_2 p_2 q_2}{}_{i_3 j_3 k_3 l_3 ,\, i_4 j_4 k_4 l_4} = & \tfrac{4}{3} \Big(  \tfrac{1}{4} \, \delta_{i_4 j_4 k_4 l_4}^{m_1 n_1 p_1 q_1} \, \delta_{i_3 j_3 k_3 l_3}^{m_2 n_2 p_2 q_2} - 2 \, \delta_{\lfloor i_4 j_4 k_4 \lceil i_3}^{m_1 n_1 p_1 q_1} \, \delta_{l_4 \rfloor j_3 k_3 l_3 \rceil}^{m_2 n_2 p_2 q_2} \\
& +2 \, \delta_{\lfloor i_4 \lceil i_3 j_3 k_3}^{m_1 n_1 p_1 q_1} \, \delta_{j_4 k_4 l_4 \rfloor l_3 \rceil}^{m_2 n_2 p_2 q_2} - \tfrac{1}{4} \, \delta_{i_3 j_3 k_3 l_3}^{m_1 n_1 p_1 q_1} \, \delta_{i_4 j_4 k_4 l_4}^{m_2 n_2 p_2 q_2} \Big) \, .
\end{align}
Define the sectional curvatures in the following way
\begin{align} \label{Definition sectional curvatures}
R \, \bbQ \, \bbQ & = R^{m_1 n_1 p_1 q_1 ,\, m_2 n_2 p_2 q_2}{}_{i_3 j_3 k_3 l_3 ,\, i_4 j_4 k_4 l_4} \, \bbQ_{m_1 n_1 p_1 q_1}{}^{i_3 j_3 k_3 l_3} \, \bbQ_{m_2 n_2 p_2 q_2}{}^{i_4 j_4 k_4 l_4} \nonumber \, ,\\
R \, \bbQ \, \bbP & = R^{m_1 n_1 p_1 q_1 ,\, m_2 n_2 p_2 q_2}{}_{i_3 j_3 k_3 l_3 ,\, i_4 j_4 k_4 l_4} \, \bbQ_{m_1 n_1 p_1 q_1}{}^{i_3 j_3 k_3 l_3} \, \bbP_{m_2 n_2 p_2 q_2}{}^{i_4 j_4 k_4 l_4} \nonumber \, ,\\
R \, \bbP \, \bbP & = R^{m_1 n_1 p_1 q_1 ,\, m_2 n_2 p_2 q_2}{}_{i_3 j_3 k_3 l_3 ,\, i_4 j_4 k_4 l_4} \, \bbP_{m_1 n_1 p_1 q_1}{}^{i_3 j_3 k_3 l_3} \, \bbP_{m_2 n_2 p_2 q_2}{}^{i_4 j_4 k_4 l_4} \, ,
\end{align}
they take the form (\ref{Sectional curvature xz}).

\providecommand{\href}[2]{#2}\begingroup\raggedright\endgroup

\end{document}